\documentclass[12pt,preprint]{aastex}
\singlespace
\slugcomment{Accepted for the publication in the {\it Astrophysical Journal}}
\def\la{\mathrel{\hbox{\rlap{\hbox{\lower4pt\hbox{$\sim$}}}\hbox{$<$}}}}
\def\ga{\mathrel{\hbox{\rlap{\hbox{\lower4pt\hbox{$\sim$}}}\hbox{$>$}}}}

\shortauthors{Park}
\shorttitle{SNR 1987A}

\begin{document}

\title{Evolutionary Status of SNR 1987A at the Age of Eighteen}

\author{Sangwook Park}

\affil{Department of Astronomy and Astrophysics,
525 Davey Lab., Pennsylvania State University, University Park, PA. 16802} 
\email{park@astro.psu.edu}

\author{Svetozar A. Zhekov\altaffilmark{1}}

\affil{JILA, University of Colorado, Box 440, Boulder, CO. 80309}

\author{David N. Burrows, Gordon P. Garmire, Judith L. Racusin}

\affil{Department of Astronomy and Astrophysics,
525 Davey Lab., Pennsylvania State University, University Park, PA. 16802}

\author{and}

\author{Richard McCray }

\affil{JILA, University of Colorado, Box 440, Boulder, CO. 80309}

\altaffiltext{1}{Current address: The Space Research Institute, Moskovska str. 6, 
Sofia-1000, Bulgaria}

\begin{abstract}

$\sim$18 yr after the supernova explosion, the blast wave of SNR 1987A 
is entering the main body of the equatorial circumstellar material, which
is causing a dramatic brightening of the remnant. We recently reported the 
observational evidence for this event from our {\it Chandra} data 
(Park et al. 2005b; P05 hereafter). We present here the temporal 
evolution of the X-ray emitting shock parameters and the detailed 
description of the spectral and image analysis of SNR 1987A, on which 
P05 was based. While the remnant becomes brighter, the softening of the 
overall X-ray spectrum continues and is enhanced on around day $\sim$6200 
(since the explosion). The two-component shock model indicates that the 
electron temperatures have been changing for the last $\sim$6 yr. 
The X-ray spectrum is now described by $kT$ $\sim$ 0.3 keV and 2.3 keV 
thermal plasmas which are believed to characteristically represent the
shock-heated density gradient along the boundary between the H{\small II}
region and the dense inner ring. As the blast wave sweeps through the 
inner circumstellar ring shining in X-rays, we expect that the shock 
parameters continue to change, revealing the density and abundance 
structure of the inner ring. Follow-up {\it Chandra} observations will 
thus uncover the past history of the progenitor's stellar evolution. 
The origin of the relatively faint hard X-ray emission ($E$ $>$ 3 keV) 
from SNR 1987A is still unclear (thermal vs. nonthermal). Considering 
the continuous brightening of the hard band intensity, as well as the 
soft band flux, follow-up monitoring observations will also be essential 
to reveal the origin of the hard X-ray emission of SNR 1987A.

\end{abstract}


\section {\label {sec:intro} INTRODUCTION}

We continue to monitor the X-ray remnant of supernova (SN) 1987A with 
the Advanced CCD Imaging Spectrometer (ACIS) on board {\it Chandra X-Ray 
Observatory}. As of 2005 July, we have performed a total of 13
observations, including a pair of deep gratings observations. 
Results from the first eight ACIS imaging observations have been 
reported in the literature \citep{bur00,park02,park04,park05a}. 
We have also presented the results from the high resolution X-ray 
spectroscopic studies of SNR 1987A using deep {\it Chandra} gratings 
observations \citep{michael02,zhekov05,zhekov06}. 

These studies revealed that the soft X-ray emission of SNR 1987A 
originates from the shocked hot gas between the forward and reverse 
shock of the blast wave, which appears as a ring-like morphology with 
the high resolution ACIS images. The X-ray emitting hot gas is primarily 
distributed in the ``disk'' containing the dense equatorial inner ring 
that was produced by asymmetric stellar winds from the massive progenitor 
star before the SN \citep{bur95,lund91,luo91}. As the shock sweeps up 
continuously increasing amount of circumstellar material (CSM), the soft 
X-ray intensity has thus been rapidly increasing. The overall X-ray 
spectrum is described by two thermal components, which 
characteristically represents the decelerated shock front ($kT$ $\sim$ 
0.2 keV) entering the dense protrusions of the inner ring and the fast 
shock ($kT$ $\sim$ 2.5 keV) propagating into less dense medium.  

The recent {\it Chandra} data indicate that the soft X-ray lightcurve of 
SNR 1987A has been even more rapidly increasing than ever since early 2004
(day $\sim$6200 since the SN). The decelerated, soft component of 
the shock ($kT$ $\sim$ 0.3 keV) begins to dominate the observed 0.5 $-$
2 keV band X-ray flux at the same epoch. The soft X-ray images also show 
a global brightening, rather than local intensity increases, during the 
same period. We interpret that these latest results are indications of 
the blast wave now interacting all the way around the inner ring rather 
than with only a few small dense protrusions primarily in the eastern side
of the remnant. We have recently reported these 
new results of the soft X-ray lightcurve and its implications 
(Park et al. 2005b, P05 hereafter). This is an exciting event that was 
predicted to occur in $\sim$10 $-$ 20 yr after the SN (e.g., 
Luo et al. 1994). The recent high resolution X-ray spectroscopy with 
the deep {\it Chandra} gratings observations consistently supports 
this new evolutionary phase of SNR 1987A \citep{zhekov05,zhekov06}. 
In this paper, we present the data analysis of all up-to-date ACIS 
observations of SNR 1987A, on which the results of P05 were based, and 
discuss how the X-ray remnant is evolving as of 2005 July ($\sim$18 yr 
after the SN).

\section{\label{sec:obs} OBSERVATIONS \& DATA REDUCTION}

All 13 {\it Chandra} observations of SNR 1987A are presented in 
Table~\ref{tbl:tab1}. The results from the gratings observations are 
presented in the literature \citep{michael02,zhekov05,zhekov06}. We 
present the results from the 11 undispersed ACIS observations in this work.
We reduce these data following the methods described in our previous
work \citep{bur00,park02}. We screened all 11 data sets with the flight 
timeline filter and turned off the pixel randomization for the highest 
possible angular resolution. We then corrected the spatial and spectral 
degradation of the ACIS data caused by radiation damage, known as 
the charge transfer inefficiency (CTI; Townsley et al. 2000), with the 
methods developed by Townsley et al. (2002a), before further 
standard data screening by status, grade, and energy selections. 
``Flaring'' pixels were removed and {\it ASCA} grades (02346) were selected. 
Photons between 0.3 keV and 8.0 keV were extracted for the data analysis. 
Lightcurves around the source region were examined for possible contamination 
from variable background emission. Overall, no severe variability was found.
With ObsID 5580, there was moderately flaring background (by a factor 
of $\sim$2) for the entire S3 chip for a short time period of $\sim$700 s. 
Although this flaring background was negligible for the small source 
region, we excluded this $\sim$700 s time interval from the data analysis.  
The pileup fraction has been small ($<$10\%) and thus was ignored 
in the analysis. However, considering the rapid brightening of SNR 1987A, 
we used subarrays for the latest 3 ACIS observations (Table~\ref{tbl:tab1})
in order to avoid potential pileup effects. 

We then applied the ``sub-pixel resolution'' method \citep{tsunemi01} 
to improve the angular resolution of the images to better than the CCD 
pixel size (0$\farcs$492). A typical improvement in the angular resolution 
by $\sim$10\% is expected from this method \citep{mori01}. The angular 
size of SNR 1987A is small (the inner ring is only about 1$\farcs$6 
across; e.g., Burrows et al. 1995; Jakobsen et al. 1991), and the ACIS 
detector pixel size is not adequate to fully resolve the remnant. 
In order to further improve the effective angular resolution 
of the ACIS images, we deconvolved the images with the detector point
spread function using a maximum likelihood algorithm \citep{rich72,lucy74} 
as described in our previous work \citep{bur00}. 

\section{\label{sec:image} X-RAY MORPHOLOGY }

\begin{figure}[]
\figurenum{1}
\centerline{\includegraphics[angle=0,width=0.9\textwidth]{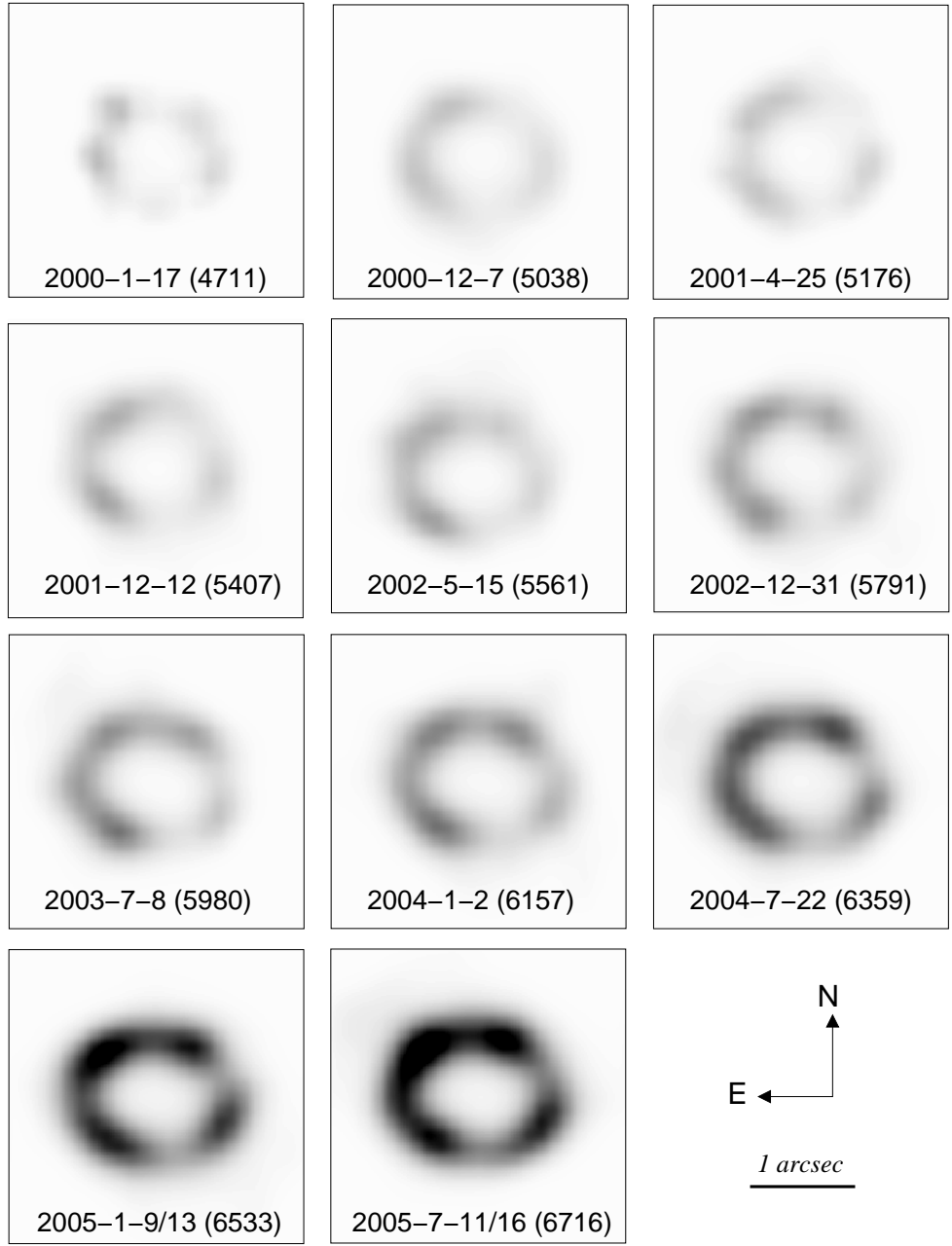}}
\figcaption[]{The 0.3$-$8.0 keV broadband images of SNR 1987A. Each 
image is exposure-corrected and the darker gray-scales correspond to 
higher intensities. The image deconvolution has been applied and 
then the images have been smoothed by convolving with a Gaussian 
($\sim$0$\farcs$1 FWHM). In each panel, the observation date and the age 
(day since the SN, in the parentheses) of the SNR are presented.
\label{fig:fig1}}
\end{figure}

\begin{figure}[]
\figurenum{2}
\centerline{\includegraphics[angle=0,width=0.5\textwidth]{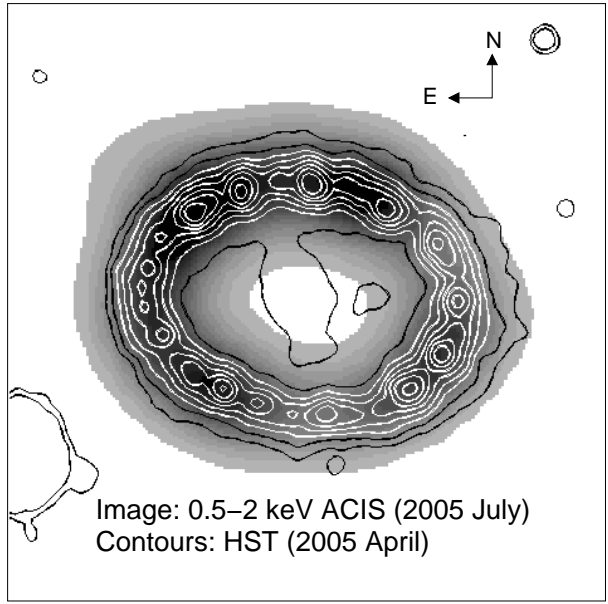}}
\figcaption[]{The soft X-ray (0.5$-$2 keV) image from the latest ACIS 
observations (taken in 2005 July) overlaid with the optical image contours 
(taken in 2005 April) from {\it the Hubble Space Telescope (HST)}. 
The unpublished {\it HST} data have been provided by Peter Challis.
\label{fig:fig2}}
\end{figure}

X-ray images of SNR 1987A from our 11 {\it Chandra}/ACIS observations are
presented in Figure~\ref{fig:fig1}. The continuous brightening 
of the X-ray emission from the SNR, particularly from the western side in
the recent images in addition to the eastern side, is evident. Our early 
{\it Chandra} observations showed that the bright soft X-ray spots 
in the eastern side of the remnant were positionally coincident 
with the bright optical spots which were then also dominating in the 
eastern parts of the inner ring \citep{bur00,park02}. Since then,
the optically bright spots have been emerging in the western parts
of the inner ring in addition to the eastern portions \citep{suger02}. 
The current X-ray images also reveal bright X-ray emission from all 
around the inner ring. These X-ray and optical data suggest that the 
blast wave shock is now engulfing areas around the entire inner ring 
several years after reaching the first local protrusions in the eastern 
side. In fact, the latest optical and X-ray bright spots show excellent 
positional correlations (Figure~\ref{fig:fig2}). As of 2005 July, 
the overall surface brightness is an order of magnitude higher than 
it was in 2000 January. A dramatic brightening of SNR 1987A, caused by 
the interaction of the blast wave with the dense inner ring, has been 
predicted to occur around 1999$-$2007 \citep{luo91,luo94,chev95,bor97}. 
More recently, based on the radial expansion of the radio remnant of 
SN 1987A, Manchester et al. (2002) predicted that this event may happen 
in 2004$\pm$2. Our {\it Chandra} images indicate that the predicted 
brightening is now underway (see P05 for more discussion). 

The radial expansion of the X-ray remnant of SN 1987A ($v$ $\sim$ 4000 $-$
5000 km s$^{-1}$) has been detected with the earlier {\it Chandra}/ACIS 
images \citep{park02,park04}. We continue to measure the radial expansion 
with the new X-ray images. While the X-ray remnant is continuously 
expanding with an average rate of $v$ $\sim$ 3200 km s$^{-1}$,
the expansion rate has evidently decelerated down to $v$ $\sim$ 1600 
km s$^{-1}$ since day $\sim$6200 (Racusin et al. 2006, in preparation).
This turnover in the expansion rate is indeed expected assuming that the 
bulk of the X-ray emission originates from the interaction between the 
blast wave and the inner ring. Because the blast wave is rapidly approaching
the dense CSM ``wall'' of the inner ring's main body, a significant
deceleration of the overall shock velocity is eventually anticipated.
The detailed description and the results from our measurements of the
radial expansion rate are presented elsewhere (Racusin et al. 2006 in
preparation).
 
The broad-subband images consequently become indistinguishable between
the soft ($E$ = 0.3 $-$ 0.8 keV) and the hard ($E$ = 1.2 $-$ 8.0 keV)
bands (Figure~\ref{fig:fig3}). This is in contrast to the early subband
images which clearly exhibited distinctive differences in bright X-ray 
spots between the soft and the hard bands \citep{park02}.  

\section{\label{sec:spectrum} SPECTRAL ANALYSIS}

Unlike the early data, which could be described by a $kT$ $\sim$ 3 keV
thermal plasma (e.g., Park et al. 2002), the recent data of SNR 1987A 
cannot be adequately fitted by a single shock model. For instance, 
although the overall spectral shape of the latest data (2005 July) may be 
described with a $kT$ $\sim$ 1.5 keV plasma, the fit is statistically 
unacceptable (${\chi}^2/{\nu}$ = 283.4/130). The poor fit is primarily 
caused by the infeasibility of the single shock model to properly fit the 
line features in the soft band ($E$ $<$ 1.5 keV). This change is most 
likely because of the complex shock structure that has recently 
developed due to the significant interaction between the blast wave and the 
ambient density gradient \citep{park04,zhekov06}. We thus perform spectral 
analysis of SNR 1987A utilizing two-component plane-parallel shock model 
\citep{bor01}, following the methods used in our previous work 
\citep{park04}. Albeit admittedly simplified, we confirm that the 
two-shock model provides a good approximation for the complex 
velocity/temperature distribution of the blast wave by characteristically 
representing the fast and decelerated shocks \citep{zhekov06}. 

We extract the source spectrum of SNR 1987A from a circular region with 
a 2$^{\prime\prime}$$-$3$^{\prime\prime}$ radius for each observation. 
The background spectrum is estimated from a surrounding annulus with an 
inner radius of 3$\farcs$5$-$4$^{\prime\prime}$ and an outer radius of 
7$\farcs$5$-$10$^{\prime\prime}$. Each spectrum has been binned to 
contain a minimum of 20$-$50 counts per channel. For the spectral 
analysis of our CTI-corrected data, we have utilized the response 
matrices appropriate for the spectral redistribution of the CCD, as 
generated by Townsley et al. (2002b). We use non-equilibrium ionization 
(NEI) shock models ({\tt vpshock} in conjuction with the NEI version 2.0 
in the {\tt XSPEC}) which are based on {\tt ATOMDB} \citep{smith01}. 
Inner-shell processes are added in this atomic database, which is 
important for the NEI plasma yet missing in the current {\tt XSPEC} NEI 
version 2\footnote{The unpublished version of the updated model has been 
provided by K. Borkowski.}. We also regenerated the ancillary response 
functions using the ACIS {\tt caldb version 3.00} for all eleven 
observations in order to consistently correct for the quantum efficiency 
degradation of the ACIS over the $\sim$6 yr period of the observations. 

The spectral contributions of the line emission from the elements 
He, C, Ca, Ar, and Ni are insignificant in the fitted energy range 
(0.4$-$5.0 keV), thus we fix the abundances of these elements at previous 
measurements. He (= 2.57, hereafter, the abundances are relative to solar 
\citep{anders89}) and C (= 0.09) are set to the abundances of the inner 
ring \citep{lund96}. We fix the Ca (= 0.34), Ar (= 0.54) and Ni (= 0.62) 
abundances to values appropriate for the LMC ISM \citep{russell92} 
because the ring abundances were unavailable for these elemental species 
in Lundqvist \& Fransson (1996). Abundances of other species, N (= 0.76), 
O (=0.09), Ne (= 0.29), Mg (= 0.24), Si (= 0.28), S (=0.45), and Fe (= 0.16), 
are fixed at values that we measured with the recent deep {\it Chandra}/Low 
Energy Transmission Gratings Spectrometer (LETG) observations 
\citep{zhekov06}, because the high resolution gratings spectrum should 
provide the most reliable measurements of the abundances. We fix the 
redshift of SNR 1987A at the value for the LMC ($v$ = 286 km s$^{-1}$; 
Michael et al. [2002]; Zhekov et al. [2005] and references therein). 

The eleven individual spectra of SNR 1987A represent a wide range in
photon statistics ($\sim$600$-$27000 counts) with 
temporal evolution of the shock parameters. In order to make reliable 
comparisons of the shock parameters among the individual observations, 
it is essential to determine the foreground absorption, which should be 
constant over the $\sim$6 yr observation period. We found that the $N_H$ 
measurement is particularly important for reliable measurements of the 
shock parameters of the early observations with low photon statistics. 
In order to make the most reliable estimate of the foreground column $N_H$, 
we fit all 11 spectra simultaneously, varying the electron temperature, 
ionization timescale, and the normalization freely among the individual 
observations. We fit $N_H$, but tie it to be the same for all individual 
epochs. The best-fit $N_H$ toward the entire SNR 1987A is 
2.35$^{+0.09}_{-0.08}$ $\times$ 10$^{21}$ cm$^{-2}$ ($\chi^2/{\nu}$ 
= 1058.5/1015; the uncertainties are with a 90\% confidence). We then 
repeat the spectral fitting for each eleven individual spectrum with $N_H$ 
and elemental abundances fixed at the values obtained above. 
We display the ACIS spectrum of SNR 1987A in two representative epochs
of 2000 January and 2005 July in Figure~\ref{fig:fig4}. The modeled 
shock parameters from these spectral fits are presented in 
Table~\ref{tbl:tab2}. The ionization timescale ($n_et$) for the soft 
component is high at all epochs and is not well constrained with lower 
limits of $n_et$ $\ga$ 10$^{12}$ cm$^{-3}$ s. The soft component $n_et$ 
is thus not explicitly presented in Table~\ref{tbl:tab2}. 

\section{\label{sec:disc} Discussion}

The X-ray morphology of SNR 1987A has been evolving from a faint, partial 
ring with a pair of relatively bright spots in the eastern side, 
to a bright, complete ring over the last $\sim$6 yr. Based on the standard 
picture of the SNR 1987A system, these morphological changes are indeed
expected as the blast wave encounters an increasing number of protrusions of
the dense inner ring and eventually sweeps through the main body of the
entire inner ring. This exciting phase of SNR 1987A's evolution
is highlighted by the recent upturn in the soft X-ray lightcurve.
The evolution of the X-ray luminosity of SNR 1987A is presented in
Table~\ref{tbl:tab3}. It is remarkable that the soft X-ray ($E$ = 
0.5 $-$ 2 keV) flux increase rate has turned up since day $\sim$6200 
(Figure~\ref{fig:fig5}, also see Figure~1 in P05). 
This soft X-ray lightcurve is interpreted as 
evidence for the blast wave reaching the entire inner ring at around 
day $\sim$6200. The detailed description and discussion of the lightcurve 
analysis can be found in P05. P05 also found observational 
evidence that independently supports this interpretation, such as the 
morphological changes in the soft X-ray intensity ratios and the changes 
in the fractional contribution from the decelerated shock for the 
observed soft X-ray flux since day $\sim$6000 $-$ 6200. Racusin et al. 
(2006, in preparation) reports an apparent deceleration of the radial 
expansion rate of the SNR since day $\sim$6200, which is self-consistent 
with our results. The low shock velocities ($v$ $<$ 1700 km s$^{-1}$) 
deduced from the line broadenings obtained with the recent deep 
{\it Chandra}/LETG observations of SNR 1987A \citep{zhekov05} were also 
supportive of the conclusions by P05. The recent mid-IR observations
revealed a similar upturn in the IR intensity at around the same
epoch of day $\sim$6000 \citep{bouchet06}. The high resolution mid-IR
images show that the bulk of IR emission originates from the inner
ring, being consistent with the soft X-ray and optical emission. 
These IR observations thus also support our interpretation of the
blast wave now interacting with the main portion of the dense inner
ring.

The X-ray spectrum can be described by two components: e.g., as of 2005
July, the soft ($kT$ $\sim$ 0.3 keV) and the hard ($kT$ $\sim$ 2.3 keV) 
components characteristically representing the decelerated and the fast 
shocks, respectively. The soft component appears to be in (or close to)
collisional ionization equilibrium (CIE) with high ionization timescales 
($n_et$ $\ga$ 10$^{12}$ cm$^{-3}$ s), while the hard component is in NEI 
condition ($n_et$ $\sim$ 2 $-$ 3 $\times$ 10$^{11}$ cm$^{-3}$ s). These 
results are consistent with our physical picture of the X-ray emission 
from the shock interacting with the dense CSM with complex density 
structure: i.e., on average, for the last several years, the soft component
represents the decelerated shock interacting with the dense CSM and the 
hard component indicates the {\it undecelerated}, fast shock propagating 
into the less dense medium. We however note that, recently, the observed
0.5 $-$ 2 keV band emission is dominated almost entirely by emission from 
the shocked dense CSM (P05). This suggests that even the fast shock component
($kT$ $\sim$ 2.3 keV) of the observed 0.5 $-$ 2 keV band X-ray emission 
now originates primarily from the ``reflected'' shock close to the inner 
ring rather than from the shocked H{\small II} region \citep{zhekov05}
(also see Figure~3 in Michael et al. 2000). 
In this picture, the decelerated shock component ($kT$ $\sim$ 0.3 keV)
of the 0.5 $-$ 2 keV flux is produced by the ``transmitted'' shock front 
into the dense inner ring. The nature of these shock components are 
discussed in detail with the high spectral resolution LETG data 
\citep{zhekov05,zhekov06}. In the current work, we focus our discussion 
on the temporal evolution of these shock components. 

The overall X-ray spectrum of SNR 1987A has been continuously softening,
particularly since day $\sim$6200 (Figure~\ref{fig:fig6}). The
significant change in the softness of the spectrum suggests that
a large portion of the shock started to interact with the dense CSM 
since day $\sim$6200. The electron temperature of the fast shock
component has been decreasing whereas that of the soft, decelerated shock 
component is increasing (Table~\ref{tbl:tab2} \& Figure~\ref{fig:fig7}). 
This ``merging'' of two characteristic temperatures may be interpreted 
as a ``transitioning'' effect that the increasing fraction of the 
low-energy tail of the fast shock emission apparently shifts into the 
hard-tail of the slow shock spectrum. While the separation between the 
two characteristic electron temperatures becomes smaller, the overall 
softening of the X-ray spectrum, as reported in the literature 
\citep{park04,park05a}, continues. As of 2005 July, the overall X-ray 
spectral shape roughly corresponds, on average, to a $kT$ $\sim$ 1.5 keV 
plasma, whereas it was $kT$ $\sim$ 3 keV in 2000 January 
\citep{bur00,park02}. This evolution of the electron temperature is 
consistent with our interpretation that a large fraction of the shock 
front has begun interacting with a significant density gradient between 
the less dense H{\small II} region and the dense inner ring. As the blast 
wave eventually sweeps through the main portions of the dense inner ring, 
the X-ray spectrum may be dominated by the density structure of the inner 
ring itself. For instance, if the inner ring has a relatively {\it uniform} 
density, one might expect that the observed X-ray spectrum would become 
properly described by a single, average electron temperature in the future. 

The volume emission measures (EM) for both shock components continue 
to significantly increase (Table~\ref{tbl:tab2}). This result is in
agreement with the increasing contribution from the reflected shock
in the fast shock component of the X-ray emission: i.e., the continuously 
increasing amount of the soft X-ray emission from the fast shock 
component is originating from the shock entering the strong density 
gradient near the boundary of the inner ring, rather than from the shocked 
tenuous H{\small II} region. As the blast wave enters a density gradient, 
the electron temperature of the shocked plasma is inversely correlated 
with the density ($T$ $\propto$ $n^{-1}$). The evolution of the electron 
temperature ratio between the fast and decelerated shocks 
(Figure~\ref{fig:fig7}) then implies changes in the density ratio from 
$n_{slow}/n_{fast}$ $\sim$ 15 (day $\sim$4700) to $\sim$7 (day $\sim$6700). 
While the EMs are increasing for both shock components, their relative
ratio has been showing no significant changes with an average ratio
of $\sim$7.3 (Figure~\ref{fig:fig8}). This means that the X-ray emitting
volume ratio has been changing from $V_{fast}/V_{slow}$ $\sim$ 30
(day $\sim$4700) to $\sim$7 (day $\sim$6700).

The density ratio between the inner ring and the H{\small II} region 
was estimated to be $\sim$18 $-$ 30 with the previous {\it Chandra}
data \citep{park04}. Considering the simple modeling and various 
geometrical assumptions embedded in the density estimates, these density 
ratios were concluded to be consistent with those ($\sim$100) derived from 
the UV data \citep{chev95,lund96}. With the significantly improved X-ray
data (both in quantity and quality) and physically more realistic
modeling than Park et al. (2004) utilized, we now consider that the 
relatively low density ratios derived with the X-ray data may represent
some actual physical implications. In other words, our {\it Chandra} 
observations presented in this work have been performed since early
2000, $\sim$3 yr after the emergence of the optical Spot 1. The emergence
of the optical Spot 1 marks the beginning of the significant interaction
of the blast wave with the dense CSM along the inner ring. Therefore, even
the earliest data of our {\it Chandra} observations would have been 
affected by the density gradient in the vicinity of the inner ring, even 
for the fast shock component. In such a case, it is not surprising to 
obtain less density contrast between the fast and the decelerated shock 
components from the X-ray data, compared with the inner ring to H{\small 
II} region density ratios measured by the UV/optical data. Nonetheless, 
without particular geometrical assumptions regarding the X-ray emitting 
regions for each component, the measured shock parameters indicate 
a small volume with a high density for the decelerated shock and
a large volume with a low density for the fast shock component.
These density and volume ratios have also been continuously reduced.
These results are in good agreement with our physical scenario 
of the soft X-ray production from the blast wave shock interaction with 
a density gradient at the boundary of the inner ring: i.e., the slow, 
transmitted shock into the small, dense regions of the inner ring and 
the fast, reflected shock off the dense CSM back into a large volume
region. 

The ionization timescale ($n_et$) of the fast shock component has been 
relatively constant for the last $\sim$6 yr (Figure~\ref{fig:fig9}). 
These $n_et$ values are in plausible agreement with the $\sim$10 $-$ 15 yr 
old shock (the SNR age since day $\sim$1200) propagating into an ambient
medium of $n_e$ $\sim$ a few 10$^2$ cm$^{-3}$ (e.g., as of 2005 July, 
the measured $n_et$ corresponds to $n_e$ $\sim$ 500 cm$^{-3}$). 
Although accurate measurements of the ionization timescales of the shocked 
plasma are difficult with the low resolution CCD spectrum, these results 
are qualitatively consistent with the overall picture of the shock-ISM 
interaction in the SNR 1987A's inner ring system and are also in good 
agreement with the results from the high resolution spectroscopy of our
{\it Chandra}/LETG observations \citep{zhekov06}.
 
\begin{figure}[]
\figurenum{3}
\centerline{\includegraphics[angle=0,width=0.9\textwidth]{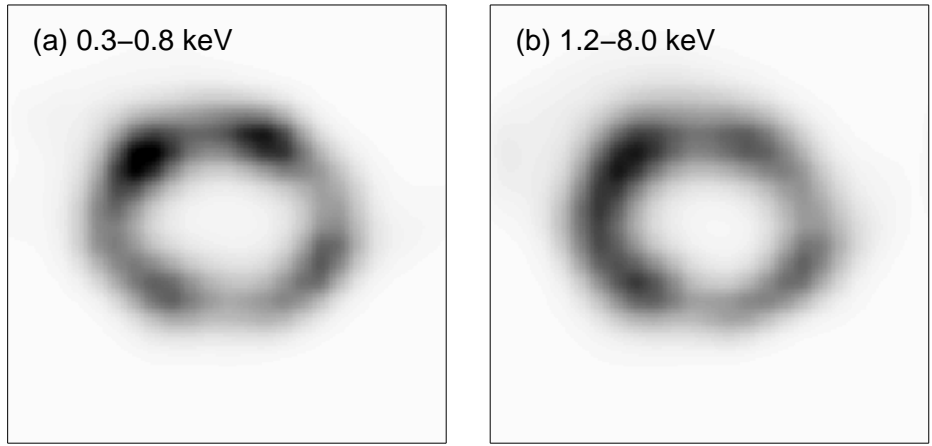}}
\figcaption[]{(a) The 0.3 $-$ 0.8 keV and (b) 1.2 $-$ 8 keV subband images 
of SNR 1987A as of 2005 July. Each image has been processed in the same way 
as those in Figure~\ref{fig:fig1}. The image orientation is also the same
as Figure~\ref{fig:fig1}.
\label{fig:fig3}}
\end{figure}

\begin{figure}[]
\figurenum{4}
\centerline{\includegraphics[angle=0,width=0.9\textwidth]{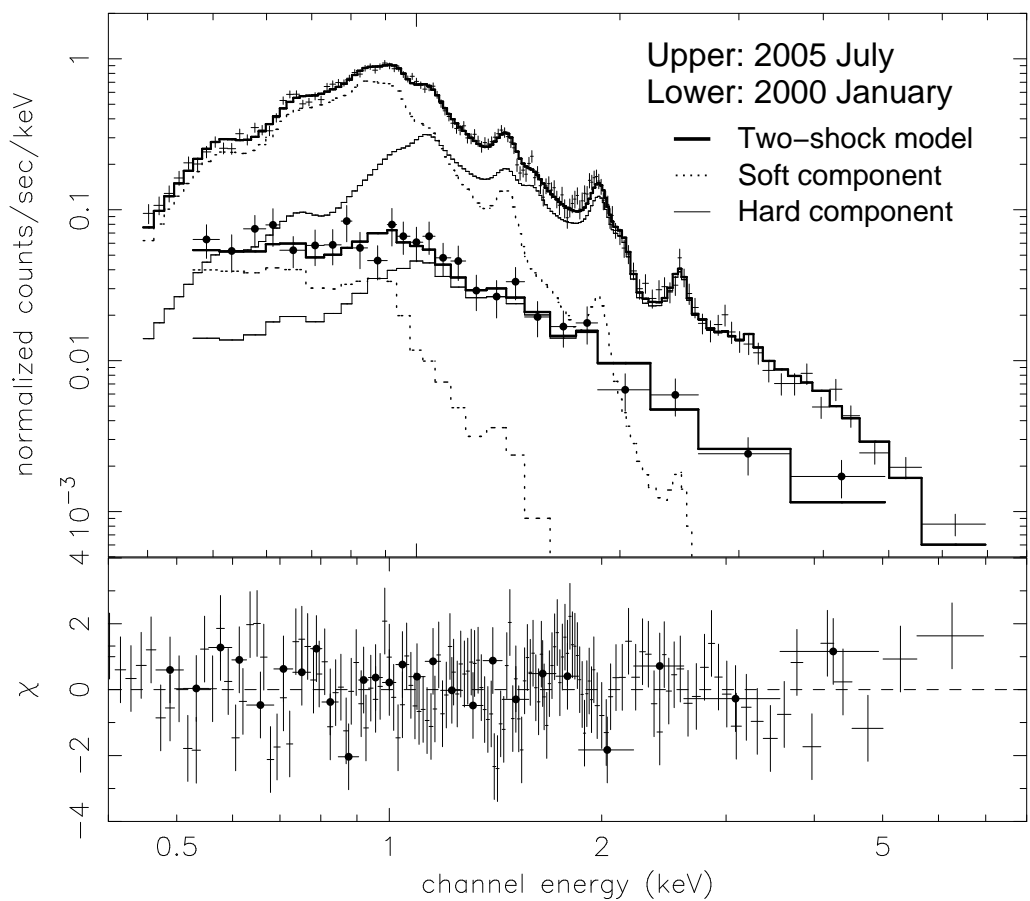}}
\figcaption[]{X-ray spectrum of SNR 1987A as obtained with the ACIS-S3.
The spectrum in two epochs are representatively displayed. The upper 
plot is the 2005-July spectrum (no markers) and the lower plot 
(filled circles) is 2000-January spectrum. The overlaid thick-solid 
lines are the best-fit two-shock model for each spectrum. For each 
spectrum, the soft and the hard sub-components of the best-fit 
two-shock model are also overlaid. The upper dotted line is the soft
component for the 2005-July spectrum, and the lower dotted line is
the soft component for 2000-January spectrum. The upper and lower 
thin-solid lines are the hard component for the 2005-July and 
the 2000-January spectrum, respectively.
\label{fig:fig4}}
\end{figure}

\begin{figure}[]
\figurenum{5}
\centerline{\includegraphics[angle=0,width=0.8\textwidth]{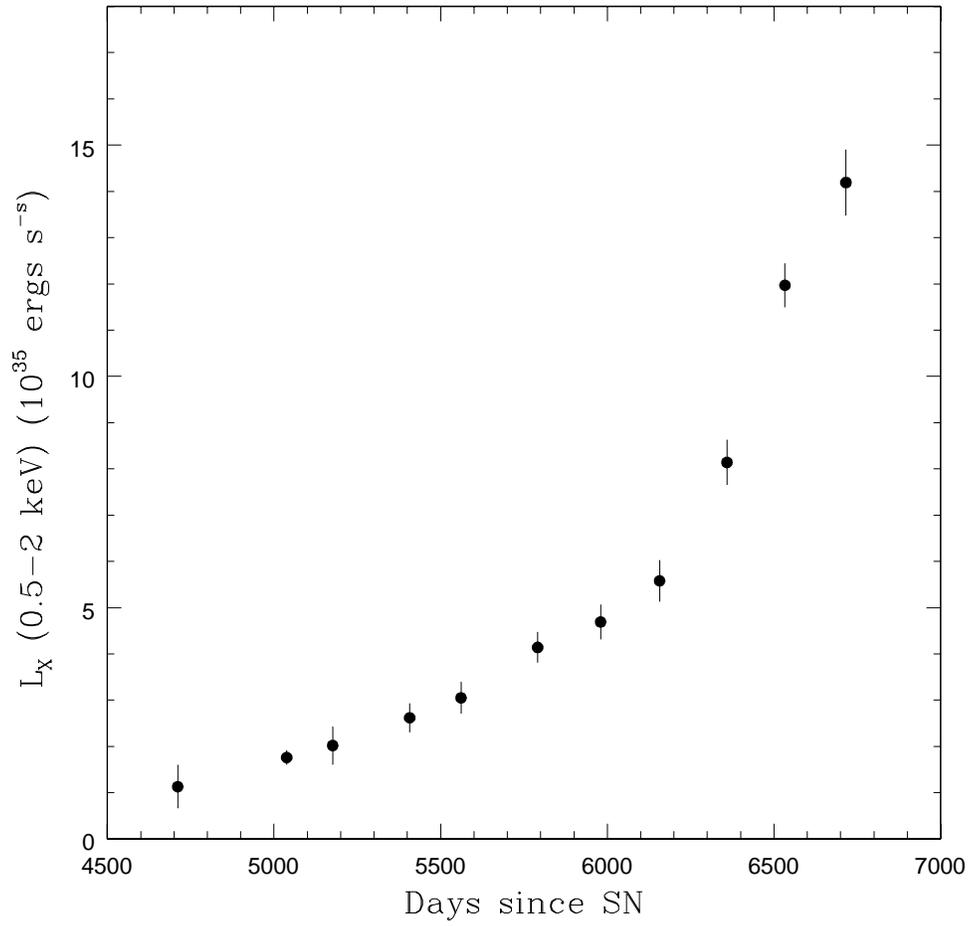}}
\figcaption[]{The 0.5 $-$ 2 keV band soft X-ray lightcurve of  
SNR 1987A as presented in Table~\ref{tbl:tab3}.
\label{fig:fig5}}
\end{figure}

\begin{figure}[]
\figurenum{6}
\centerline{\includegraphics[angle=0,width=0.5\textwidth]{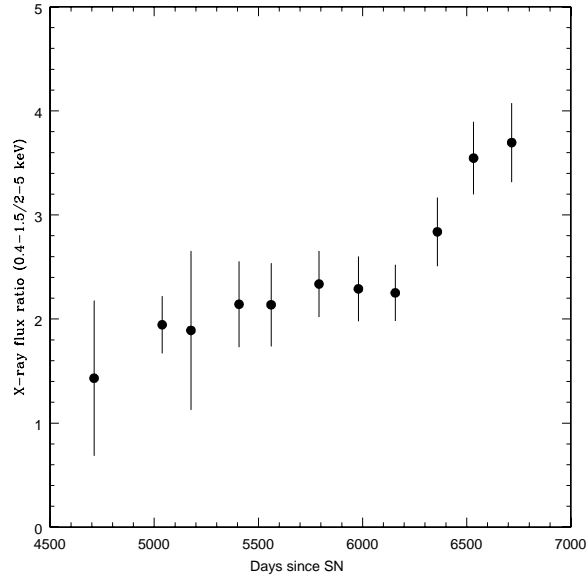}}
\figcaption[]{The observed 0.4$-$1.5 to 2$-$5 keV band X-ray flux ratio 
of SNR 1987A.
\label{fig:fig6}}
\end{figure}

\begin{figure}[]
\figurenum{7}
\centerline{\includegraphics[angle=0,width=0.5\textwidth]{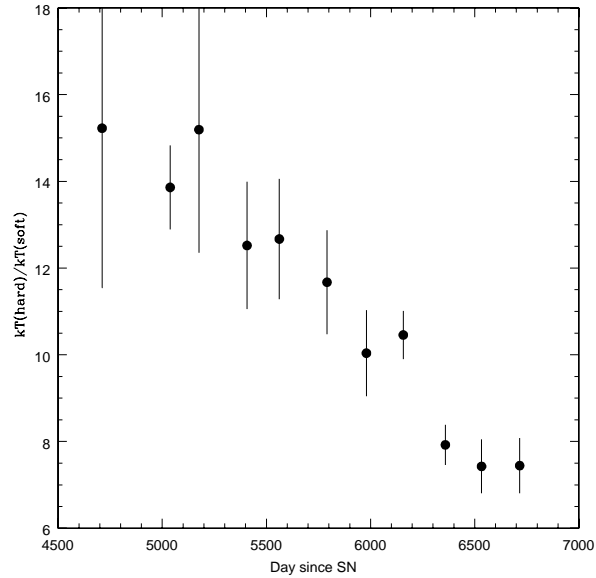}}
\figcaption[]{The electron temperature ratios between the hard and the 
soft component shocks from SNR 1987A.
\label{fig:fig7}}
\end{figure}

\begin{figure}[]
\figurenum{8}
\centerline{\includegraphics[angle=0,width=0.8\textwidth]{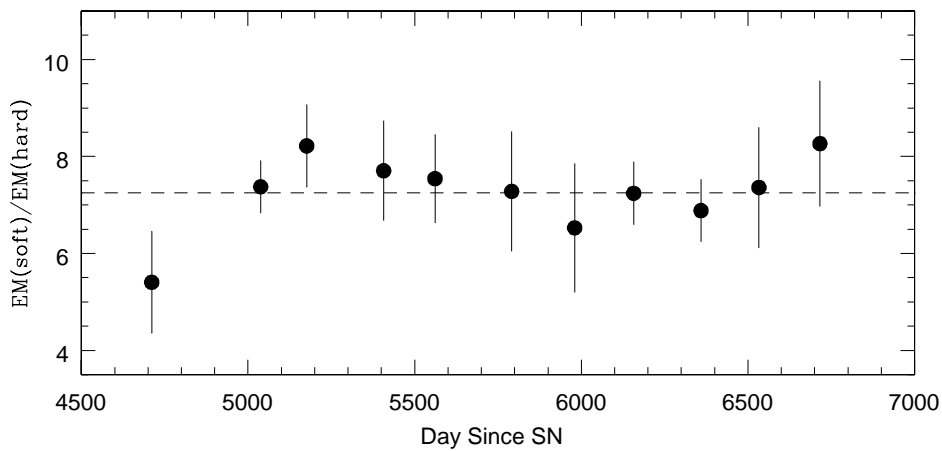}}
\figcaption[]{The emission measure ratios between the soft and the 
hard component shocks from SNR 1987A. The horizontal dashed line represents
the average ratio.
\label{fig:fig8}}
\end{figure}

\begin{figure}[]
\figurenum{9}
\centerline{\includegraphics[angle=0,width=0.8\textwidth]{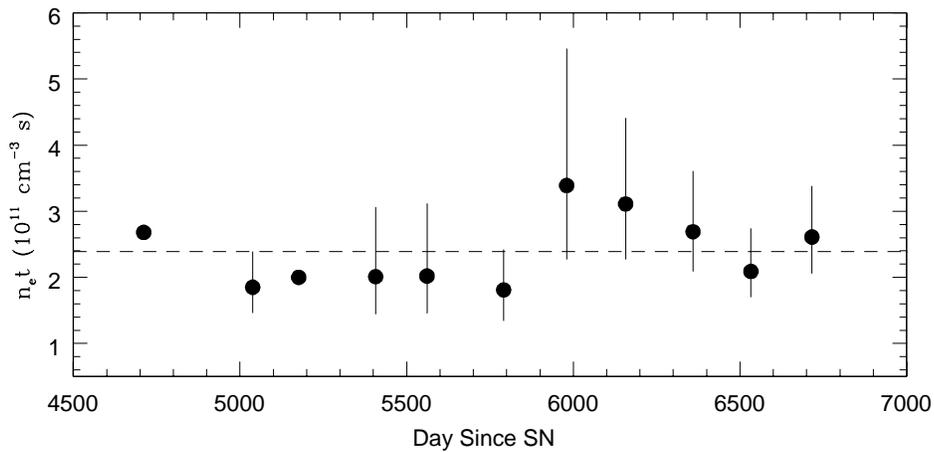}}
\figcaption[]{The ionization timescale ($n_et$) of the hard component
of SNR 1987A spectrum. The horizontal dashed line represents
the average $n_et$ value. Note that $n_et$ for two epochs (day 4711 and 5176)
is unconstrained because of the poor photon statistics. Only the ``best-fit''
values are presented for them without error bars.
\label{fig:fig9}}
\end{figure}

On the other hand, the increase rate of the hard band intensity ($E$ = 
3 $-$ 10 keV) is significantly lower than that of the soft X-rays ($E$ = 
0.5 $-$ 2 keV) (Table~\ref{tbl:tab3}). This lower increase rate of the 
hard X-ray emission might result simply from the continuous softening of 
the overall X-ray spectrum. The similar morphology between the soft and
the hard band images is supportive of this interpretation 
(Figure~\ref{fig:fig3}, see also Figure~4 in P05). P05, however, noticed
that the hard X-ray increase rate appears to be consistent with that
of radio lightcurves. A nonthermal origin for the hard X-ray emission
from the synchrotron radiation in the particle acceleration sites thus
cannot be ruled out (P05). On the other hand, an additional power law 
component is not required to adequately fit the observed X-ray spectrum. 
The presence of the hard power law component therefore cannot be 
determined with the current data because of the low photon statistics 
in the hard band. The true origin of the hard X-ray emission from 
SNR 1987A is thus currently uncertain. Although fainter than the soft 
band emission, the hard X-ray intensity is steadily increasing. 
Follow-up monitoring observations in X-rays and radio will therefore be 
useful to reveal the nature of the hard X-ray emission in coming years.

\acknowledgments

The authors thank P. Challis and the Supernova INtensive Study (SINS)
collaboration for providing their {\it HST} images. We also thank
K. Borkowski for providing us the updated NEI models.
This work was supported in part by the Smithsonian Astrophysical
Observatory under {\it Chandra} grant GO5-6073X.

\clearpage

\begin{deluxetable}{ccccc}
\footnotesize
\tablecaption{Chandra Observations of SNR 1987A.
\label{tbl:tab1}}
\tablewidth{0pt}
\tablehead{\colhead{Observation ID} & \colhead{Date (Age\tablenotemark{a})} & 
\colhead{Instrument} & \colhead{Exposure (ks)} & \colhead{Counts}}
\startdata
      124+1387\tablenotemark{b} & 1999-10-6 (4609) & ACIS-S + HETG & 116.1 & 
690\tablenotemark{c} \\
      122 & 2000-1-17 (4711) & ACIS-S3 & 8.6 & 607 \\
      1967 & 2000-12-07 (5038) & ACIS-S3 & 98.8 & 9030 \\
      1044 & 2001-4-25 (5176) & ACIS-S3 & 17.8 & 1800 \\
      2831 & 2001-12-12 (5407) & ACIS-S3 & 49.4 & 6226 \\
      2832 & 2002-5-15 (5561) & ACIS-S3 & 44.3 & 6427 \\
      3829 & 2002-12-31 (5791) & ACIS-S3 & 49.0 & 9277 \\
      3830 & 2003-7-8 (5980) & ACIS-S3 & 45.3 & 9668 \\
      4614 & 2004-1-2 (6157) & ACIS-S3 & 46.5 & 11856 \\
      4615 & 2004-7-22 (6359) & ACIS-S3 (1/2 subarray) & 48.8 & 17979 \\
      4640+4641+5362 & 2004-8-26$\sim$9-5 & ACIS-S + LETG & 289.0 & 
16557\tablenotemark{c} \\
      +5363+6099\tablenotemark{b} & ($\sim$6400) & & & \\
      5579+6178\tablenotemark{b} & 2005-1-12 (6533) & ACIS-S3 (1/8 subarray) & 
48.3 & 24939 \\
      5580+6345\tablenotemark{b} & 2005-7-14 (6716) & ACIS-S3 (1/8 subarray) & 
44.1 & 27048 \\
\enddata
\tablenotetext{a}{Day since SN.}
\tablenotetext{b}{These observations were splitted by multiple sequences 
which were combined for the analysis.}
\tablenotetext{c}{Photon statistics are from the zeroth-order data.}
\end{deluxetable}

\begin{deluxetable}{lcccccc}
\footnotesize
\tablecaption{Best-fit Parameters from the Two-Shock Model Fit of 
SNR 1987A\tablenotemark{a}
\label{tbl:tab2}}
\tablewidth{0pt}
\tablehead{\colhead{Age\tablenotemark{b}} & \colhead{$kT(soft)$\tablenotemark{c}} & 
\colhead{$kT(hard)$\tablenotemark{c}} & \colhead{$n_et(hard)$\tablenotemark{c}} &
\colhead{$EM(soft)$\tablenotemark{c}} & \colhead{$EM(hard)$\tablenotemark{c}} &
\colhead{$\chi^2$/$\nu$} \\
\colhead{(Days)} & \colhead{(keV)} & \colhead{(keV)} & \colhead{(10$^{11}$ cm$^{-3}$ s)} 
& \colhead{(10$^{57}$ cm$^{-3}$)} & \colhead{(10$^{57}$ cm$^{-3}$)} & }
\startdata 
\vspace{1.0mm}
4711\tablenotemark{d} & 0.222$^{+0.032}_{-0.029}$ & 3.38$^{+2.82}_{-1.20}$ &
    2.68 (fixed) & 24.00$^{+6.00}_{-6.30}$ & 4.44$^{+0.87}_{-0.66}$ & 19.0/21 \\
\vspace{1.0mm}
5038 & 0.228$^{+0.011}_{-0.006}$ & 3.16$^{+0.49}_{-0.33}$ & 1.85$^{+0.54}_{-0.39}$ 
     & 39.60$^{+3.00}_{-5.40}$ & 5.37$^{+0.45}_{-0.39}$ & 96.0/98 \\
\vspace{1.0mm}
5176\tablenotemark{d} & 0.235$^{+0.017}_{-0.015}$ & 3.57$^{+1.33}_{-0.96}$ & 
     2.00 (fixed) & 47.09$^{+7.20}_{-6.00}$ & 5.73$^{+0.72}_{-0.54}$ & 22.8/41 \\
\vspace{1.0mm}
5407 & 0.238$^{+0.056}_{-0.008}$ & 2.98$^{+0.74}_{-0.46}$ & 2.01$^{+1.05}_{-0.57}$
     & 58.49$^{+4.80}_{-26.40}$ & 7.59$^{+0.99}_{-0.87}$ & 78.4/76 \\
\vspace{1.0mm}
5561 & 0.251$^{+0.049}_{-0.009}$ & 3.18$^{+0.63}_{-0.62}$ & 2.02$^{+1.09}_{-0.56}$
     & 64.49$^{+7.80}_{-26.70}$ & 8.55$^{+1.17}_{-0.78}$ & 62.8/78 \\
\vspace{1.0mm}
5791 & 0.269$^{+0.045}_{-0.029}$ & 3.14$^{+0.68}_{-0.34}$ & 1.81$^{+0.61}_{-0.47}$
     & 77.09$^{+26.10}_{-24.00}$ & 10.59$^{+0.99}_{-1.14}$ & 92.8/100 \\
\vspace{1.0mm}
5980 & 0.272$^{+0.078}_{-0.017}$ & 2.73$^{+0.38}_{-0.28}$ & 3.39$^{+2.07}_{-1.12}$
     & 89.69$^{+21.30}_{-39.00}$ & 13.74$^{+1.47}_{-1.44}$ & 122.2/101 \\
\vspace{1.0mm}
6157 & 0.263$^{+0.021}_{-0.009}$ & 2.75$^{+0.33}_{-0.22}$ & 3.11$^{+1.30}_{-0.84}$
     & 121.18$^{+15.00}_{-23.40}$ & 16.74$^{+1.50}_{-1.44}$ & 132.4/109 \\
\vspace{1.0mm}
6359 & 0.284$^{+0.044}_{-0.010}$ & 2.25$^{+0.20}_{-0.19}$ & 2.69$^{+0.92}_{-0.60}$
     & 160.48$^{+25.20}_{-26.10}$ & 23.31$^{+1.65}_{-1.83}$ & 167.7/123 \\
\vspace{1.0mm}
6533 & 0.303$^{+0.065}_{-0.025}$ & 2.25$^{+0.19}_{-0.20}$ & 2.09$^{+0.65}_{-0.39}$
     & 202.47$^{+58.49}_{-67.49}$ & 27.51$^{+2.28}_{-1.68}$ & 109.0/133 \\
\vspace{1.0mm}
6716 & 0.309$^{+0.051}_{-0.025}$ & 2.30$^{+0.20}_{-0.17}$ & 2.61$^{+0.77}_{-0.55}$
     & 255.87$^{+63.59}_{-68.39}$ & 30.96$^{+2.64}_{-2.16}$ & 151.9/135 \\
\enddata

\tablenotetext{a}{$N_H$ is fixed at 2.35 $\times$ 10$^{21}$ cm$^{-2}$.}
\tablenotetext{b}{Days since the SN.}
\tablenotetext{c}{2$\sigma$ uncertainties.}
\tablenotetext{d}{The errors are estimated with $n_et$ parameters fixed
at the best-fit values.}

\end{deluxetable}

\begin{deluxetable}{lccc}
\footnotesize
\tablecaption{ X-Ray Luminosity of SNR 1987A\tablenotemark{a}
\label{tbl:tab3}}
\tablewidth{0pt}
\tablehead{\colhead{Age\tablenotemark{b}} & \colhead{$L_X$ (0.5$-$2.0 keV)} & 
\colhead{$L_X$ (3$-$10 keV)} & \colhead{$L_X$ (0.5$-$10 keV)}\\
\colhead{(Days)} & \colhead{(10$^{35}$ ergs s$^{-1}$)} & \colhead{(10$^{35}$ ergs 
s$^{-1}$)} & \colhead{(10$^{35}$ ergs s$^{-1}$)} 
}
\startdata 
\vspace{1.0mm}
4711 & 1.13 & 0.26 & 1.54 \\
\vspace{1.0mm}
5038 & 1.76 & 0.28 & 2.22 \\
\vspace{1.0mm}
5176 & 2.02 & 0.37 & 2.59 \\
\vspace{1.0mm}
5407 & 2.62 & 0.37 & 3.24 \\
\vspace{1.0mm}
5561 & 3.05 & 0.45 & 3.79 \\
\vspace{1.0mm}
5791 & 4.14 & 0.55 & 5.05 \\
\vspace{1.0mm}
5980 & 4.69 & 0.59 & 5.71 \\
\vspace{1.0mm}
6157 & 5.58 & 0.73 & 6.82 \\
\vspace{1.0mm}
6359 & 8.14 & 0.76 & 9.54 \\
\vspace{1.0mm}
6533 & 11.97 & 0.85 & 13.58 \\
\vspace{1.0mm}
6716 & 14.19 & 0.99 & 16.06 \\
\enddata

\tablenotetext{a}{After corrected for $N_H$ = 2.35 $\times$ 10$^{21}$ cm$^{-2}$.}
\tablenotetext{b}{Days since the SN.}

\end{deluxetable}

\end{document}